# Direct-Current Generator Based on Moving Van der Waals Schottky Diode

*Shisheng Lin[1,2,]\*, Yanghua Lu[1], Sirui Feng[1], Zhenzhen Hao[1] and Yanfei Yan[1]*

1: College of Microelectronics, College of Information Science and Electronic Engineering, Zhejiang University, Hangzhou, 310027, China

2: State Key Laboratory of Modern Optical Instrumentation, Zhejiang University, Hangzhou, 310027, China

*Corresponding author. Tel: +86-0571-87951555

Email: shishenglin@zju.edu.cn.

## Abstract

Traditionally, Schottky diodes are used statically in the electronic information industry but dynamic state Schottky diodes based applications have been rarely explored. Herein, a novel Schottky diode named "moving Schottky diode generator" has been designed, which can convert mechanical energy into electrical energy with voltage output as high as 0.6V, by means of lateral movement between graphene/metal film and semiconductor, where the semiconductor can be non-piezoelectric materials. The mechanism is based on the built-in electric field separation of drifting electrons in moving van der Waals Schottky diode. The power output can be further increased in future through optimizing the Schottky diode. The graphene film/silicon moving van der Waals Schottky diode based generator behaves better stability. This direct-current generator has the potential of converting mechanical efficiently and vibrational energy into electricity and enables many promising applications.

**One Sentence Summary:** A promising direct-current generator through moving van der Waals Schottky formed between graphene/metal film and semiconductor.

## Main

Since the invention of Schottky diode at 1947 (*1, 2*), many important applications have been developed including high power and high speed electronic devices, etc (*3-5*). However, the moving Schottky diode is rarely explored since all the applications are focused on the static structure (*6*). In 2006, Wang et al. discovered that a rectified ZnO Schottky diode can output electricity due to the combination of piezoelectric and semiconductor properties of ZnO (*7*). The varied strain on ZnO nanowire can induce the variation of the barrier height of the Schottky barrier, leading to charge and discharge of the device (*7*). Due to the requirement of the piezoelectric materials, the application of nanogenerator is limited (*8-11*). Herein, we have demonstrated a direct-current generator through a moving van der Waals Schottky diode by dynamically moving graphene or metal film over Si or GaAs substrates. The mechanism of the moving Schottky diode generator is based on the built-in electric field separation of the drifting electrons emitted by the disappearance of the depletion layer in moving Schottky diode. The working lifetime can be properly improved by choosing graphene as the metal material, which is outstanding not only because of its high conductive but also the high mechanical physical properties (*12-17*). Actually, static graphene/semiconductor van der Waals heterostructure based solar cells (*18-22*), self-driven photodetectors (*23-25*) and water flow nanogenerator (*26-29*) have been proposed. Compared with metal, graphene film has the unique advantage of high flexibility and durability, which promises the graphene film/silicon moving Schottky diode generator working well after 10000 runs back and forth. This direct-current generator has the potential of converting mechanical, vibrational and energy into electricity and enables many promising applications.

As a representative moving Schottky diode generator, the 3D diagram and schematic structure of the moving graphene film/Si Schottky diode based generator are illustrated in Figure 1A. The graphene or metal film was pressed to the semiconductor substrate compactly with a constant force when we move it on the semiconductor surface. When the graphene or metal film was dragged along the surface of the semiconductor substrate, a voltage or current output can be observed. As a representative system, the graphene film/Si heterostructure shows excellent rectification behavior with a low leakage current of ~2 nA under bias voltage of 1V (The top inset of Figure 1A). The work function of N-type Si and P-type Si we used is approximate to its conduction band or valence band, i.e. 4.30eV and 4.90eV (The calculation is listed in the Supporting Information), respectively. The work function of graphene (~4.6eV) is different from that of Si (*14*), thus a built-in electric field will form between the graphene film and P-type/N-type Si substrate. For example, when graphene film touches N-type Si substrate by van der Waals force, electrons of the side of N-type Si will inject into graphene and an electron depletion region will form in N-type Si substrate. The barrier height of graphene film/Si heterojunction can be extracted by fitting dark current density-voltage (J-V) curves, which can be written as follows:

$$J = J_0 (exp \frac{qV}{N_{IF}KT} - 1) \qquad (1)$$

where $K$ is the Boltzmann constant, $T$ is the temperature and $N_{IF}$ is the ideality factor. According to the thermionic-emission theory, $J_0$ is the reverse saturation current density, which can be expressed by:

$$J_0 = A^* T^2 \exp\left(-\frac{q\Phi_{barrier}}{KT}\right) \qquad (2)$$

where $A^*$ is the effective Richardson constant of n-Si, $\Phi_{barrier}$ is the barrier height of junction, $q$ is the quantity of electric charge. Based on Equation (1) and Equation (2), the calculated $N_{IF}$ and $\Phi_{barrier}$ of graphene/Si heterostructure is 15.2 and 0.718eV, respectively (The calculation is listed in the Supporting Information). Primarily, the graphene film was dragged along the surface of the N-type Si at a dragging speed of 2 cm/s and a voltage output up to 0.1V has been observed, which increases as the dragging speed increases (Figure 1B). As the dragging speed increases up to 4 cm/s, a voltage output of 0.18V can be achieved. The voltage output can reach a maximum value of 0.22V and remain stable when the dragging speed is as high as 6cm/s. Furthermore, the dependence of the output of moving graphene film/Si Schottky diode under the force exerted on graphene is shown in Figure 1C and it is found that a force of 5N is an optimal choice, where the diode has the best rectification characteristic (as shown in J-V curves and $N_{IF}$ of heterojunction under different force in the Figure S1). With the force increasing, the voltage output increases and then decreases with the change of the rectification characteristic of graphene/Si heterojunction. As shown in Figure 1D, some other materials such as indium tin oxid (ITO) and aluminum have been studied instead of the graphene film. The voltage output of smooth aluminum film can reach the highest voltage up to 0.6V at a dragging speed of 6 cm/s, which is three times of moving graphene film/Si Schottky diode based generator. The work function of metallic aluminum is about 4.28eV, which is smaller than Fermi energy of Si substrate, so the generated voltage direction is opposite to the graphene film. The current output of different materials also increases as the dragging speed increases, which is up to 4μA of moving aluminum film/N-type Si Schottky diode when the dragging speed reaches 6cm/s (Figure 1E). Besides, the current output also increases linearly as the work area increases, as shown in Figure 1F. In order to study the effect of semiconductor, N-type GaAs substrate has been used to measure the voltage and current output of the moving Schottky diode based generator with different speeds and materials, whose results are listed in the Figure S2 and S3. Some detailed results about N-type Si substrate based generator are also listed in the Figure S4 for comparison. Normally, Si substrate based generator has a better performance compared with GaAs substrate based generator. Overall, a novel moving Schottky diode based generator composed by a series of materials has been realized, demonstrating the repeatability and universality of the moving Schottky diode generator.

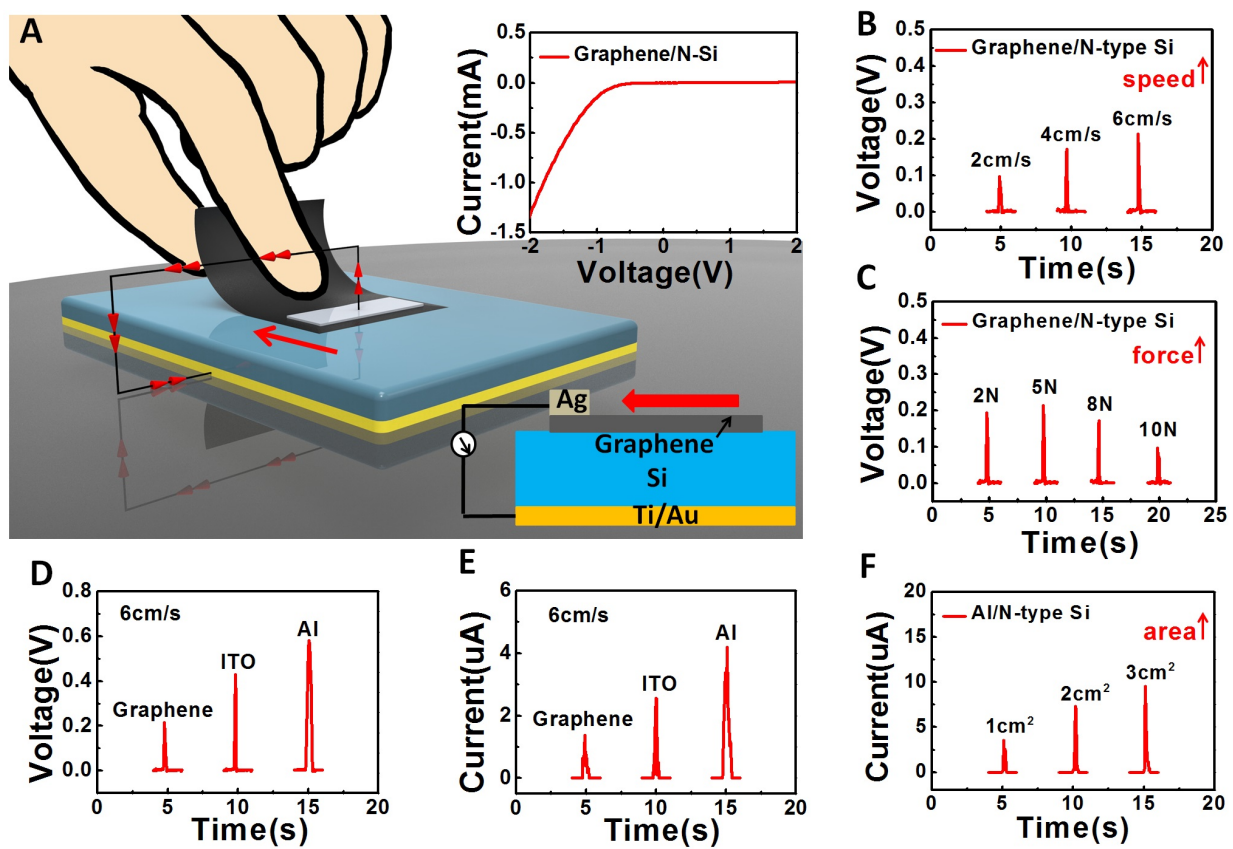

**Fig. 1. Experimental designs and results of the moving heterostructure based generator. (A) Experimental design for converting mechanical energy into electrical energy by a moving Schottky diode. Inset at top shows the J-V curve of graphene film/Si heterostructure based generator. (B) The voltage output of graphene/N-Si generator with different speeds. (C)**

**The voltage output of graphene/N-type Si generator with different forces applied to the heterojunction. (D) The voltage output of graphene, ITO, Al/N-type Si generator. (E) The current output of graphene, ITO, Al/N-type Si generator. (F) The current output of Al/N-type Si generator with different work areas.**

We propose that the dynamic appearance and disappearance of depletion layer in moving Schottky diode, which causes the separation of drifted electrons and holes, should be the origin of the electrical power generation. For the film/Si heterostructure system, the drift-diffusion equation and current can be described as below:

$$J_n = J_n^{drift} + J_n^{diff} = -qnu_n \nabla V + qD_n \nabla n \tag{3}$$

$$J_p = J_p^{drift} + J_p^{diff} = -qpu_p \nabla V - qD_p \nabla p \tag{4}$$

where $J_n$, $J_p$, $\mu_n$, $\mu_p$, $D_n$, $D_p$ are the electron/hole current density, the electron/hole mobility and the electron/hole diffusion coefficient in semiconductor, respectively. $V$ is the potential, $q$ is the elementary charge, $n$ and $p$ are the position dependent electron density and hole density in semiconductor, respectively. As schematically shown in Figure 2, the surface of metal or graphene film is rough, there are some gaps between metal or graphene film and semiconductor substrate. There is a process of appearance and disappearance of the depletion layer when the metal/graphene films move along the silicon surface. When the metal/graphene film touches with the Si substrate, the depletion layer will form at the interface of heterojunction immediately. Furthermore, as the effect of gaps in the Schottky diode, the sliding of film on the large area silicon surface will leads to the disappearance of the depletion layer and the re-establishment of the depletion layer dynamically. According to equation (3) and (4), current signal is consisted of diffusion current ($J^{diff}$) and drift current ($J^{drift}$). We assume the total contact area between the film and silicon surface stays unchanged, so the $J^{diff}$ remain unchanged. However, the dynamic appearance and disappearance of the depletion layer will generate additional drifting electrons and holes, which can be separated by the built-in electronic field, leading to the increase of $J^{drift}$. Hence, the $J^{drift}$ is larger than $J^{diff}$ in the moving Schottky diode. This built-in electric field separation of the drifting electrons and holes emitted by the appearance and disappearance of the depletion layer in heterojunction causes the generation of current signal, which is defined as the moving Schottky diode generator. When we increase the moving speed, more drifting electrons can be generated, which increases the electricity output of the generator (Figure 1B). Besides, the direction of the voltage output is different because of the different work function of materials composing the heterojunction. When we use N-type Si substrate with the work function of 4.3eV, the electrons in the metal film whose work function locates above N-type Si, such as Al, will transfer to the N-type Si while forming the built-in electric field. The moving Schottky diode induced drifting electrons will transfer to the metal film layer under built-in electric field and form electrical signal (Figure 2A and 2B). On the contrary, with the moving of heterojunction, electrons in N-type Si substrate will transfer against the metal film whose work function locates below N-type Si (Figure 2C and 2D), such as Cu. Similarly, when we use P-type Si substrate with the work function of 4.9eV, the electrons of the metal film whose work function locates above P-type Si, such as Cu, will transfer to the P-type Si while forming the built-in electric field. The moving Schottky diode induced drifting holes in P-type Si will transfer against metal film and form electrical signal (Figure 2E and 2F). On the contrary, with the moving of heterojunction, holes in P-type Si substrate will transfer to the metal film whose work function locates below P-type Si under built-in electric field (Figure 2G and 2H), such as Au. Some other cases about metal film/Si heterojunction based generator are listed in the Figure S5, which provide further evidence to support abovementioned mechanism.

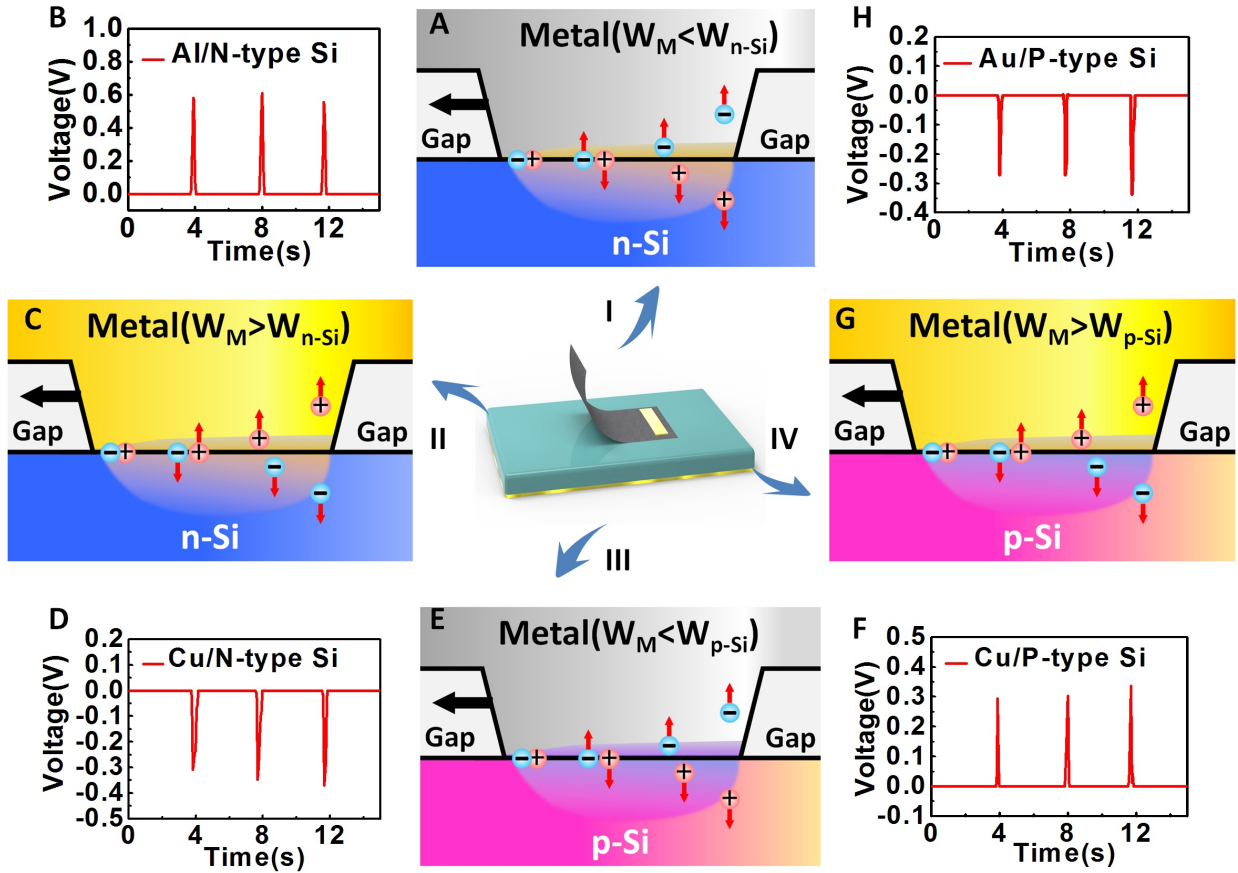

**Figure 2** The schematic diagram of the moving Schottky diode based metal film/Si heterostructure generator with (A) $W_M < W_{n-Si}$, (B) such as Al/N-type Si generator; (C) $W_M > W_{n-Si}$, (D) such as Cu/N-type Si generator; (E) $W_M < W_{p-Si}$, (F) such as Cu/P-type Si generator; (G) $W_M > W_{p-Si}$, (H) such as Au/P-type Si generator.

Furthermore, as this process is caused by the separation of moving Schottky diode induced drifting electrons and holes in Schottky diode, the generator can output continuous and direct electricity if we circularly move the film along the surface of the Si substrate. When the lateral movement between the film and semiconductor is rapid and continual, the induced voltage can be generated constantly. We have designed a generator system for continuously moving the graphene/metal film over the Si substrate (Figure S6). Indeed, a direct voltage/current output of 0.6V/4μA has been achieved (Figure 3A, 3B) and the power output of this Al/P-type Si generator is about 0.02W/m$^2$. However, the contact between the metal film and semiconductor is very tough thus induces the irreversible damage of the device, leading to the performance decline (Figure 3C). On the other hand, the graphene film can be stably used for the moving Schottky diode generator, which does not show obvious degradation after 10000 times of running whereas the metal film degrades after 1000 times of running (Figure 3D). The flexibility characteristic of graphene film ensures the smooth voltage output and limited substrate damage (Figure S7). The rectification characteristic of the graphene film/Si heterostructure has also been measured, which behaves limited damages after 10000 times of running (Figure S8). The calculated $N_{IF}$ and $\Phi_{barrier}$ of graphene/Si heterostructure after 10000 times of running is 21.2 and 0.652eV respectively, which show limited difference compared with initial values of 15.2 and 0.718eV. After analyzing the influence of the force exerted on the graphene film/Si heterojunction, a force of 5N was found to have the best rectification characteristic and maximum voltage output with limited damages to the Si substrate under continuous work condition (Figure S9). Finally, a moving graphene/Si Schottky diode based generator with great possibility for practical application is achieved, which can transfer the mechanical energy into direct current constantly. This mechanical movement process reveals a physical picture of a moving Schottky diode with the separation of drifting electrons and holes in Schottky diode.

In order to demonstrate the potential applications of this moving Schottky diode based generator, we use capacitors to collect the electricity then light up a light emitting diode (LED).

Figure 3E shows the circuit diagram of directly and continuously charging capacitors without a rectifier bridge then light up the red LED. There are 5 generators and 5 capacitors used to storage enough power to provide the voltage output for a LED, which can work in 1.8-2.4V driving voltage. Each of generators can provide nearly 0.4V voltage output, so 5 generators can light up the red LED here. The red LED has been successfully lighten after five 470μF capacitors being charged for 15 mins (Figure 3F). We can increase the output by optimizing the design of the generator structure in the future and then optimize this LED system. This direct-current generator has the potential of converting mechanical, vibrational energy into electricity and enables many promising applications.

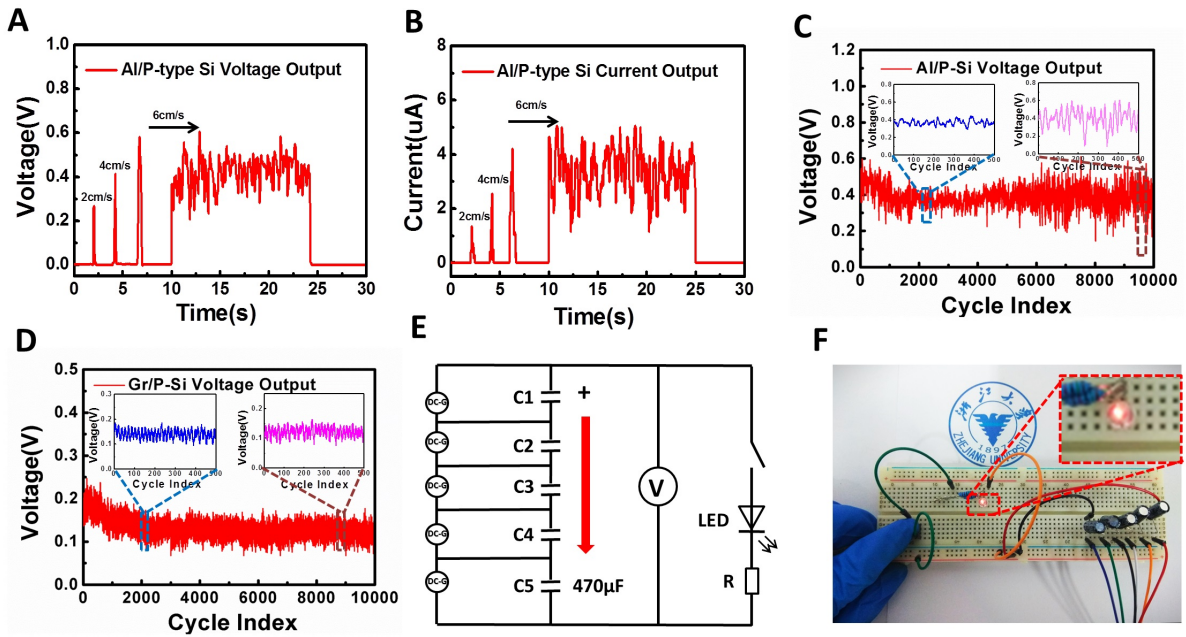

**Figure 3 The direct continuous voltage output of heterostructure based generator and circuit designed for lighting LED. (A) The direct and continuous voltage output of generator. (B) The direct and continuous current output of generator. (C) The direct and continuous voltage output of the moving graphene film/p-Si generator for 10000 cycles. (D) The direct and continuous voltage output of the moving Al film/p-Si generator for 10000**

**cycles. (E) The circuit diagram of directly and continuously charging capacitors without a rectifier bridge then light up the LED. (F) Pictures taken from video to show the luminance of a LED powered by our heterojunction based generator.**

In this work, we have converted mechanical energy into electrical energy by means of lateral movement between graphene/metal film and semiconductor. The rectifying characteristic of the Schottky barrier formed between the graphene/metal film and the semiconductor when they contacted. The mechanism highlights the built-in electric field separation of the drifting electrons emitted by the disappearance of the depletion layer in the moving Schottky diode. Lateral movement induced voltage/current of the heterojunction is as high as 0.6V and 0.04A/m$^2$. The power density of the heterostructure based generator is measured to be 0.02W/m$^2$, which can be used to light a red LED after storage by capacitor. According to the calculation listed in the Supporting Information, the power density of this moving Schottky diode based generator can reach as high as ~55300W/m$^2$, which is limited by the quality of the Schottky diode and defect in the interface. This direct-current generator has the potential of converting mechanical, vibrational and tidal energy into electricity and enables many promising applications.

**ACKNOWLEDGMENTS**

**Funding:** S. S. Lin thanks the support from the National Natural Science Foundation of China (No. 51202216, 51502264 and 61774135) and Special Foundation of Young Professor of Zhejiang University (Grant No. 2013QNA5007).

**Author Contributions:** S. S. Lin designed the experiments, analyzed the data and conceived all the works. Y. H. Lu and S. R. Feng carried out the experiments, discussed the results and wrote the paper. Z. Z. Hao and Y. F. Yan discussed the results. All authors contributed to the writing of the paper.

**Competing interests**: The authors declare no competing financial interest.

**Data and materials availability:** Supplementary materials contain additional data. All data needed to evaluate the conclusions in the paper are present in the paper or the supplementary materials. Reprints and permission information is available online. Readers are welcome to comment on the online version of the paper.